\documentclass[prd,aps,nofootinbib,onecolumn,showpacs,10pt]{revtex4-1}
\usepackage{graphicx,epsfig}

\usepackage{epsf}
\usepackage{graphicx}

\newcommand{\al}{\alpha}
\newcommand{\bt}{\beta}
\newcommand{\la}{\lambda}
\newcommand{\si}{\sigma}

\begin{document}

\title{  \bf   Scalar  Quarkonia at Finite Temperature}
\author{E. V. Veliev$^{\dag1}$, H. Sundu $^{\dag2}$, K. Azizi$^{\ddag3}$,  M. Bayar $^{\dag4}$ \\
$^{\dag}$Department of Physics , Kocaeli University, 41380 Izmit,
Turkey\\
$^{\ddag}$Physics Division,  Faculty of Arts and Sciences, Do\u gu\c s
University,
 Ac{\i}badem-Kad{\i}k\"oy, \\ 34722 Istanbul, Turkey\\
$^1$email:elsen@kocaeli.edu.tr\\
$^2$email:hayriye.sundu@kocaeli.edu.tr\\
$^3$ e-mail:kazizi@dogus.edu.tr\\
$^4$e-mail:melahat.bayar@kocaeli.edu.tr}

\begin{abstract}

Masses and decay constants of the scalar quarkonia, $\chi_{Q0} (Q=b,c)$  with quantum numbers $I^G(J^{PC})=0^{+}(0^{++})$ are calculated in the framework of the QCD sum rules approach both in
vacuum and finite temperature.  The masses and decay constants remain unchanged up to $T\simeq100~MeV$
 but they start to diminish with increasing the
temperature after this point. At near the  critic or deconfinement temperature, the decay constants reach approximately to  25\% of their values in vacuum, while the masses are decreased about 6\% and 23\% for bottom and
charm cases, respectively.
The results  at zero temperature  are in a good consistency with the existing  experimental values and predictions of the other nonperturbative approaches. Our predictions on the decay constants in vacuum as well as  the behavior of the masses and decay
constants with respect to the temperature can be checked in the future experiments.

\end{abstract}

\maketitle

\section{Introduction}
To understand the internal structure of scalar mesons has been a prominent
topic in the last 30-40 years. Although the scalar mesons have been investigated for several
decades, many properties of them are not so clear yet and
identifying the scalar mesons is difficult,  experimentally. Hence, the theoretical works can play a crucial role in this respect. In particle physics, the quarkonia refers to  flavorless mesons containing a heavy b (c) quark and its own antiquark, i.e., $b\bar b$ (bottomonium) and $c\bar c$ (charmonium). These approximately non-relativistic
systems are the best candidates to investigate the hadronic dynamics and  study the perturbative and non-perturbative aspects of QCD. It was believed that the quarkonia can help us  to extract
the nature of quark-antiquark interaction at the hadronic scale and play the same  role in probing the QCD
as the hydrogen atom play in  the atomic physics \cite{Novikov}.

A large number of the beauty and charmed systems have been experimentally observed in the last few decades (see for instance \cite{Aubert,Augustin,Choi}) and the theoretical calculations on the properties of
these systems have been made mainly using  potential model, where the  quarkonia is described by a static potential, $V=-\frac{4}{3}\frac{\alpha_s}{r}+kr$ and its extensions like the Coulomb gauge model
\cite{Ebert,Crater,Wang1,Dudek,Guo}.
The first term in the potential is related to one gluon exchange and the second term is called the confinement potential. The recent CLEO measurements  on the two-photon decay rates of the even-parity,
 P-wave scalar  $0^{++}, \chi_{b(c)0}$ and tensor $2^{++}, \chi_{b(c)2}$ states (\cite{Ecklund,CAmsler} and references therein) were motivation to investigate the properties of the quarkonia and their radiative decays
 from the   quark-antiquark interaction point of view (see for example \cite{Lansberg1,Lansberg2}).

In \cite{Luchab}, which is a recent study on extraction of ground-state decay constant from
both  sum rules and  potential models, it is stated that  results obtained at each step of the extraction procedure both in
QCD and in potential models follow the same pattern, hence all our findings concerning the extraction of bound-state parameters from correlation functions obtained in potential model can apply also to QCD.
It is also proven that in  QCD sum rules approach by tunning the continuum threshold which is related to the energy of the first exited state specially
with a Borel-parameter dependent threshold, we can get a more reliable and accurate determination of bound-state characteristics comparing the potential models. The QCD sum rules approach as a non-perturbative
 approach is one of the most powerful and applicable tools to spectroscopy of hadrons and can play a crucial role in investigation of the properties of
 the hadrons\cite{MAS,colangelo,braun,balitsky}. It has been used to
calculate the masses and decay constants of mesons
\cite{AIVainshtein,LJReinders,SNarison,
MJamin,AAPenin,Du,Kazem1,Kazem2}. This approach was extended  to
contain the properties of the hadrons  at finite temperature called
thermal QCD sum rules \cite{Bochkarev,C.Adami,T.Hatsuda} supposing
that the operator product expansion (OPE) and the quark-hadron
duality assumption remain valid, however the quark-quark,
quark-gluon and gluon-gluon
 condensates are altered by their thermal versions. The main aspiration of this addendum was to explain the results obtained from the heavy ion collision experiments.
It is presently believed that the hot and
dense medium where the hadrons are formed modifies masses and decay
widths of hadrons. It is shown that heavy mesons like $J/\psi$
and also radial and orbital $c\overline{c}$ excitations  have different
behavior when the temperature of the medium changes (see \cite{HTDing}  and references therein).
In \cite{Colangelo},  scalar mesons and scalar glueballs are
investigated in holographic QCD  at finite temperature . A flood of papers have also  been dedicated
 generally to  the determination of the condensates,  mass and decay constant of mesons and some properties of the nucleons at finite temperature
\cite{Miller1,Furnstahl,Koike,Huang,Fetea,S.Mallik,Mukherjee,Mallik,Zschocke,Dominguez0,Aliev1,Meyer,Veliev,Panero}.

In the present  work, we  calculate  the mass and decay constant of the heavy scalar  $\chi_{Q0}$  mesons with quantum numbers $I^G(J^{PC})=0^{+}(0^{++})$  using the
 thermal QCD sum rules approach.   Here, we assume that with replacing the vacuum
 condensates and also the continuum threshold by their thermal version, the  sum rules for the observables (masses and decay constants ) remain valid. In calculations, we  take into account  the
 additional operators in the Wilson expansion at finite temperature \cite{Shuryak} and modify  spectral density in QCD side. These  operators are due to
  the breakdown of Lorentz invariance at finite temperature by the selection
of the thermal rest frame, where matter is at rest at a definite temperature \cite{Mukherjee,Weldon}.  In this condition, the residual O(3) invariance
brings  these  extra operators with the same mass dimension as the vacuum condensates. We also consider   the interaction of the currents with the existing particles in the medium at finite temperature.
 Such interactions  require modification of  the hadron spectral density.

 The outline of the paper is as follows: in
 next section, sum rules for the the mass and the decay constant of the heavy
scalar,  $\chi_{Q0}$  mesons are obtained in the framework of the  QCD sum rules at  finite temperature.
Section III encompasses our numerical predictions for the mass and
 decay constants as well as comparison of the results with the existing predictions of the other non-perturbative approaches and experimental values.

\section{QCD Sum Rules for the Mass and Decay constant}
In this section, we obtain sum rules for the mass as well as the decay constant of the scalar quarkonia
containing b or c quark in the framework of
the thermal QCD sum rules.   For this aim, we will
 evaluate the two-point thermal correlation function,
\begin{eqnarray}\label{correl.func.1}
\Pi(q,T) =i\int d^{4}xe^{iq. x}{\langle} {\cal T}\left ( J^S (x) \bar J^S(0)\right){\rangle},
\end{eqnarray}
in two different ways:  physical and theoretical representations. In
correlation function $T$ denotes the temperature,  ${\cal T}$  is
the time ordering product and
$J^S(x)=\overline{Q}(x)Q(x)$ is the interpolating current of the
heavy scalar meson, $S=\chi_{Q0} (Q=b,c)$. The thermal average of
any operator, \textit{A} can be expressed as:
\begin{eqnarray}\label{A}
{\langle}A {\rangle}=\frac{Tr(e^{-\beta H} A)}{Tr( e^{-\beta H})},
\end{eqnarray}
where $H$ is the QCD Hamiltonian, and $\beta = 1/T$ is the inverse
of the temperature $T$
and traces are carried out over any complete set of states.

The physical or
phenomenological representation of the aforementioned two-point
correlation function is obtained in terms of the hadronic parameters
 saturating it with a tower of scalar mesons with the same quantum numbers
 as the interpolating current.  The  theoretical or QCD  representation is gained via operator product expansion (OPE) in terms of the QCD parameters such as quark's masses,  and the vacuum condensates considering the internal
 structure of these mesons, i.e., quarks, gluons and their
interactions with each other as well as with the QCD vacuum.
 Sum rules for the physical observables such as the  decay constant and mass are obtained equating these two different representations through dispersion relation. To suppress the
contribution of the higher states and continuum,  Borel
transformation with respect to the $Q_0^2=-q_0^2$ is applied to both sides of the sum rules for
physical quantities.

To calculate the phenomenological part, we  insert a complete
set of intermediate  states owing the same quantum numbers with current
 $J^S$ between the currents in Eq.
(\ref{correl.func.1}) and perform the integral over.  As a result, at
 $T=0$, we obtain
\begin{eqnarray}\label{phen1}
\Pi(q,0)=\frac{{\langle}0\mid J(0) \mid S\rangle \langle S \mid J(0)\mid
 0\rangle}{m_{S}^2-q^2}
&+& \cdots,
\end{eqnarray}
 where $\cdots$ represents the contributions of the  higher states and continuum and $m_{S}$ is  mass of the heavy scalar meson.
The matrix element creating the scalar meson from the vacuum can be written in terms of the   decay constant, $f_{S}$
by the following manner:
\begin{eqnarray}\label{lep}
\langle 0 \mid J(0)\mid S\rangle=f_{S}  m_{S}.
\end{eqnarray}
Note that Eqs. (\ref{phen1}) and (\ref{lep}) are valid also at finite temperature,  hence, the final
representation for the physical side can be written in terms of the temperature dependent mass and decay constant as:
\begin{eqnarray}\label{phen2}
\Pi(q,T)=\frac{f_{S}^2(T) m_{S}^2(T)}{m_{S}^2(T)-q^2} &+& \cdots
\end{eqnarray}

\begin{figure}[h!]
\begin{center}
\includegraphics[width=12cm]{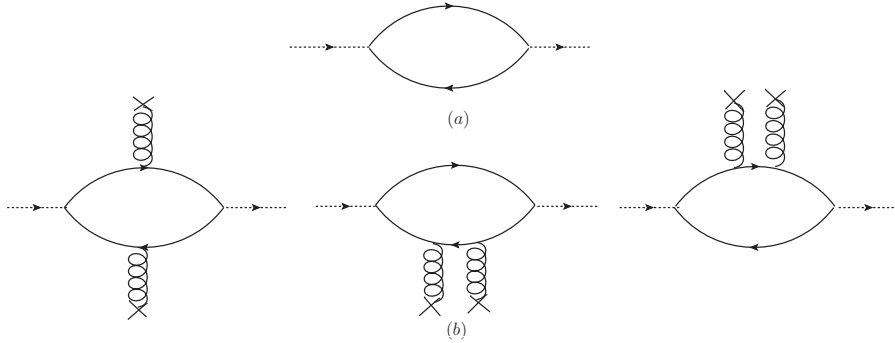}
\end{center}
\caption{ (a): Bare loop diagram
 (b): Diagrams corresponding to  gluon condensates.} \label{fig1}
\end{figure}

 In QCD side, the correlation function is calculated in deep Euclidean region, $q^2\ll-\Lambda_{QCD}^2$ via OPE where the short or perturbative and long distance or non-perturbative effects are separated, i.e.,
\begin{eqnarray}\label{correl.func.QCD1}
\Pi^{QCD}(q,T) =\Pi^{pert}(q,T)+\Pi^{nonpert}(q,T).
\end{eqnarray}
The short distance contribution  (bare loop diagram  in figure
(\ref{fig1}) part (a)) is  calculated using the perturbation theory,
 whereas the long distance   contributions (diagrams shown in figure
(\ref{fig1}) part  (b) ) are represented in terms of the thermal expectation values of some operators. To proceed, we write the perturbative part in terms of a dispersion
integral,
\begin{eqnarray}\label{correl.func.QCD1}
\Pi^{QCD}(q,T) =\int \frac{ds \rho(s,T)}{s-q^2}+\Pi^{nonpert},
\end{eqnarray}
where, $\rho(s,T)$  is called the spectral density at finite temperature.  The thermal spectral density at fixed $\mid \textbf{q}\mid$ can be expressed as:
\begin{eqnarray}\label{rhoq}
\rho(q,T)=\frac{1}{\pi}~Im\Pi^{pert}(q,T)~\tanh\left(\frac{\beta
q_{0}}{2}\right).
\end{eqnarray}
 To proceed, we need to know the fermion propagator at finite temperature. The thermal fermion  propagator at real time is given as:
\begin{eqnarray}\label{prop}
S(k)=(\gamma_{\mu}~k^{\mu}+m_{Q})\left(\frac{i}{k^2-m_{Q}^2+i\varepsilon}-2~\pi
~n~(|k_{0}|)~\delta(k^2-m_{Q}^2)\right),
\end{eqnarray}
here, $n(x)$ is Fermi distribution function,
\begin{eqnarray}\label{nf}
n(x)=\left[ exp (\beta x)+1 \right ]^{-1}.
\end{eqnarray}
Using the above propagator, we find the following expression for the imaginary part of the correlation function  at $\mid \textbf{q}\mid=0$ limit:
\begin{eqnarray}\label{ImPhi}
Im\Pi(q_0,T)=N_{c}\int\frac{d\textbf{k}}{8\pi^2}
\frac{1}{\omega^2} (q_{0}~ \omega -2
m_{Q}^2)\Big(1-2~n(\omega)+2~n^2(\omega)\Big)\delta(q_{0}-2~\omega),
\end{eqnarray}
where, $\omega=\sqrt{m_Q^2+\textbf{k}^2}$. After some straightforward calculations, the thermal
spectral density is obtained as:
\begin{eqnarray}\label{rhoper}
\rho(s)=\frac{3s}{8
\pi^2}\left(1-\frac{4m_Q^2}{s}\right)^{\frac{3}{2}}
\left(1-2~n\left(\frac{\sqrt{s}}{2}\right)\right).
\end{eqnarray}
In the non-perturbative part, the main  contribution comes from the two
gluon condensates since the heavy quark condensates are suppressed
by inverse powers of the heavy quark mass and can be safely removed.
The gluon condensate diagrams are represented in part (b) of
figure (\ref{fig1}). In order to calculate nonperturbative
contributions, we use Fock-Schwinger
gauge,  $x^{\mu}A^{a}_{\mu}(x)=0$. In momentum space, the  vacuum gluon field is expressed as:
\begin{eqnarray}\label{Amu}
A^{a}_{\mu}(k')=-\frac{i}{2}(2 \pi)^4 G^{a}_{\rho
\mu}(0)\frac{\partial} {\partial k'_{\rho}}\delta^{(4)}(k'),
\end{eqnarray}
and in calculations, we use the  quark-gluon-quark vertex as:
\begin{eqnarray}\label{qgqver}
\Gamma_{ij\mu}^a=ig\gamma_\mu
\left(\frac{\lambda^{a}}{2}\right)_{ij},
\end{eqnarray}
where $k'$ is the gluon momentum.

  After straightforward
calculations, the non-perturbative part in momentum space is obtained as:
\begin{eqnarray}\label{npPi}
&&\Pi^{nonpert}\nonumber\\
&&=\int^{1}_{0} dx\frac{~x^2}{288\pi(m_Q^2+q^2
~x~(-1+x))^4}\Big\{3\langle \alpha_s
G^2\rangle\Big[40m_Q^6x^2-9q^6x^2(-1+x)^4(-1-2x+2x^2)-12m_Q^2q^4x(-1+x)^2
\nonumber\\
&&\times(1+4x-12x^2+6x^3)+m_Q^4q^2(-15+156x-441x^2+434x^3-134x^4)
\Big]-\alpha_s \langle
u^{\alpha}\Theta^{g}_{\alpha\beta}u^{\beta}\rangle\Big[
4q^2(-1+x)\Big(q^4(-1+x)^2
\nonumber\\
&&\times x^2(9+11x-14x^2+12x^3)+m_Q^4(-15+135x-246x^2+176x^3)
+4m_Q^2q^2x(-3-8x+28x^2-34x^3+17x^4)\Big)
\nonumber\\
&&-16(-1+x)(q.u)^2\Big(q^4x^2(-1+x)^2(9+11x-14x^2+12x^3)+m_Q^4
(-15+135x-246x^2+176x^3)
\nonumber\\
&& +4m_Q^2q^2x(-3-8x+28x^2-34x^3+17x^4)\Big) \Big]\Big\},
\end{eqnarray}
where,  four-vector $u^{\mu}$  is  the velocity of the heat bath and it is introduced to restore Lorentz invariance formally in the thermal field theory.
In the rest  frame of the heat bath,  $ u^{\mu} = (1, 0, 0, 0)$ and $u^2 = 1$. In deriving  the above expression, we have used the following relation considering the Lorentz covariance
 \cite{S.Mallik}:
\begin{eqnarray}\label{}
\langle Tr^c G_{\al \bt} G_{\la \si}\rangle  = (g_{\al \la} g_{\bt \si} -g_{\al \si}
g_{\bt \la})A  -(u_{\al} u_{\la} g_{\bt \si} -u_{\al}
u_{\si} g_{\bt \la}
                             -u_{\bt} u_{\la} g_{\al \si} +u_{\bt} u_{\si}
g_{\al \la})B .
    \end{eqnarray}
Contracting indices on both sides, we obtain,
\begin{eqnarray}
A &=& {1\over 24} \langle G^a_{\al \bt} G^{a \al \bt}\rangle
+{1\over 6}\langle u^{\al} {\Theta}^g_{\al \bt} u^{\bt}\rangle, \\
B &=& {1\over 3}\langle u^{\al} {\Theta}^g _{\al \bt} u^{\bt}\rangle,
\end{eqnarray}
where,  $\Theta^{g}_{\alpha\beta}$ is   the traceless, gluonic part of the stress-tensor of the QCD  and
it is defined as:
\begin{eqnarray}\label{}
\Theta^{g}_{\alpha\beta}=-G_{\alpha\lambda}^{a}G_{\beta}^{\lambda a}+\frac{1}{4}g_{\alpha\beta}G_{\lambda\sigma}^{a}G^{\lambda\sigma a}.
\end{eqnarray}

Matching the phenomenological and QCD sides of the correlation
function, the sum rules  for the mass and decay constant  of  scalar
meson are obtained. To suppress the contribution of the higher
states and continuum,  Borel transformation over  $q^2$ as
well as continuum subtraction are performed. As a result of the
above procedure, we obtain the following sum rule
 for the decay constant:
\begin{eqnarray}\label{lepsum}
m_S^2(T)f_S^2(T)e^{\frac{-m_S^2(T)}{M^2}}=\left\{
\int_{4m_Q^2}^{s_{0}(T)}
ds~\rho(s)~e^{-\frac{s}{M^{2}}}+\hat{B}\Pi^{nonpert}\right\},
\end{eqnarray}
where $M^2$ is the Borel mass parameter and $s_{0}(T)$ is the temperature dependent continuum threshold. The sum rules for the mass is obtained
applying derivative with respect to $-\frac{1}{M^2}$  to the both
sides of the sum rule for the  decay constant of the scalar meson in
Eq. (\ref{lepsum}) and dividing by itself:
\begin{eqnarray}\label{mass2}
m_S^2(T)=\frac{\int_{4m_Q^2}^{s_{0}(T)}
ds~s~\rho(s)~exp(-\frac{s}{M^{2}})+\Pi^{nonpert}_{1}(M^2,T)}{\int_{4m_Q^2}^{s_{0}(T)}
ds~\rho(s)~exp(-\frac{s}{M^{2}})+\hat{B}\Pi^{nonpert}(M^2,T)},
\end{eqnarray}
where
\begin{eqnarray}\label{Pinp1}
\Pi^{nonpert}_{1}(M^2,T)=- \frac{d}{d(1/M^2)}\hat{B}\Pi^{nonpert}(M^2,T),
\end{eqnarray}
and
$\hat{B}\Pi^{nonpert}(M^2,T)$ shows  contribution of the gluon condensates in
Borel transformed scheme and it is given by:
\begin{eqnarray}\label{rhononper}
\hat{B}\Pi^{nonpert} &=&\int^{1}_{0} dx  e^{\frac{m_{Q}^2 }{M^2
x(x-1)}}\frac{1}{96 M^6 \pi (x-1)^4
x^3}\left\{\vphantom{\int_0^{x_2}}\left[\vphantom{\int_0^{x_2}}
\langle \alpha_s G^2\rangle \left(\vphantom{\int_0^{x_2}}m_Q^6(1-2
x)^2 (-3+5 x)
\right.\right.\right.\nonumber\\
&+&9 M^6 (-1+x)^4 x^3(-1-2 x+2 x^2)-3 m_Q^2 M^4 (-1+x)^2 x^2 (-5+7
x-12 x^2+6 x^3)
\nonumber\\
&+&2m_Q^4 M^2 x (3-21x+48x^2-41x^3+11x^4)\left.
\vphantom{\int_0^{x_2}}\right) -4\alpha_s \langle
\Theta^g\rangle
\left(\vphantom{\int_0^{x_2}}m_Q^6(1-2x)^2(-3+5x)\right.
\nonumber\\
&+& M^6(-1+x)^3x^3(9+11x-14x^2+12x^3)+2m_Q^4 M^2
x(3-23x+58x^2-57x^3+19x^4)
\nonumber\\
&-&m_Q^2~ M^4~ (-1+x)^2 x^2 (-15+11x-26x^2+32x^3) \left.
\vphantom{\int_0^{x_2}} \right) \left.
\vphantom{\int_0^{x_2}}\right] \left.\vphantom{\int_0^{x_2}}\right\}.
\end{eqnarray}
 We use the gluonic part of energy density both obtained from lattice QCD \cite{MCheng} and chiral perturbation theory \cite{P.Gerber}. In the rest  frame of the heat bath, the results obtained in  \cite{MCheng}
at
 lattice QCD are reproduced well by the following fit parametrization for the thermal average of total energy density, $\langle \Theta \rangle$,
\begin{eqnarray}\label{tetag}
\langle \Theta \rangle= 2 \langle \Theta^{g}\rangle= 6\times10^{-6}exp[80(T-0.1)](GeV^4),
\end{eqnarray}
where temperature $T $ is measured in units of $GeV$ and this parametrization is valid in the interval
 $0.1~GeV\leq T \leq 0.17~GeV$. Note that the total energy density has been known for $T\geq0 $ in the chiral perturbation theory, while this quantity has only been calculated
 for $T\geq100MeV$ in lattice QCD (see \cite{Miller1} and
 \cite{MCheng}).
In  low temperature chiral perturbation limit, the results presented in \cite{P.Gerber} are better described by the expression,
\begin{eqnarray}\label{tetagchiral}
\langle \Theta\rangle= \langle
\Theta^{\mu}_{\mu}\rangle +3~p,
\end{eqnarray}
where,     $p$ is pressure and $\langle
\Theta^{\mu}_{\mu}\rangle$ is trace of the total energy momentum tensor. They are given as:
\begin{eqnarray}\label{tetamumu}
\langle
\Theta^{\mu}_{\mu}\rangle=\frac{\pi^2}{270}\frac{T^{8}}{F_{\pi}^{4}}
\ln \Big[\frac{\Lambda_{p}}{T}\Big], ~~~~~~~~~~~~~~p=
3~T~\Big(\frac{m_{\pi}~T}{2~\pi}\Big)^{\frac{3}{2}}\Big(1+\frac{15~T}{8~m_{\pi}}+\frac{105~T^{2}}{128~
m_{\pi}^{2}}\Big)exp\Big[-\frac{m_{\pi}}{T}\Big],
\end{eqnarray}
where $\Lambda_{p}=0.275GeV$, $F_{\pi}=0.093GeV$ and
$m_{\pi}=0.14GeV$.

Our final task in this section is to introduce the temperature
dependent continuum threshold, $s_0(T)$, gluon condensate, $\langle
G^2\rangle$ and the strong coupling constant.
 The temperature dependent    continuum threshold \cite{CADominguez} and gluon condensate \cite{Miller1,MCheng} are well
described by the following fit parameterizations:
\begin{eqnarray}\label{sT}
s(T)= s_{0}\left[\vphantom{\int_0^{x_2}}
1-\Big(\frac{T}{T^{*}_{c}}\Big)^8\vphantom{\int_0^{x_2}}\right]+4~m_Q^2~
\left(\vphantom{\int_0^{x_2}}\frac{T}{T^{*}_{c}}
\vphantom{\int_0^{x_2}} \right)^8,
\end{eqnarray}
\begin{eqnarray}\label{G2TLattice}
\langle G^2\rangle_=\frac{\langle
0|G^2|0\rangle}{exp\left[\vphantom{\int_0^{x_2}}12\Big(\frac{T}{T_{c}}-1.05\Big)
\vphantom{\int_0^{x_2}}\right]+1},
\end{eqnarray}
where $T^{*}_{c}=1.1\times T_c=0.176GeV$, and $s_0$ and
$\langle0|G^2|0\rangle$  are the continuum threshold and   the gluon
condensate in vacuum, respectively. These parameterizations are valid
only in the interval $0~ \leq T \leq 170~MeV$. Here, we should stress that the continuum threshold presented above is equal to the continuum threshold in vacuum at
$T=0$ but it starts to diminish increasing the temperature such that at $T=T^{*}_c$ it reaches the perturbative QCD threshold,  $4m_Q^2$. This parametrization belongs to
heavy-heavy system and differ considerably with the case of light-light and heavy-light quark systems, where the continuum threshold
is related to the thermal light quark condensate (for details see \cite{CADominguez}).

We also use temperature dependent strong coupling constant \cite{Kaczmarek,SuHoungLee} as:
\begin{eqnarray}\label{geks2T}
g^{-2}(T)=\frac{11}{8\pi^2}\ln\Big(\frac{2\pi
T}{\Lambda_{\overline{MS}}}\Big)+\frac{51}{88\pi^2}\ln\Big[2\ln\Big(\frac{2\pi
T}{\Lambda_{\overline{MS}}}\Big)\Big],
\end{eqnarray}
where, $\Lambda_{\overline{MS}}\simeq T_c/1.14$ and in numerical calculations, instead of the $\alpha_s$ in front of $\langle \Theta^{g}\rangle$ in Eq. (\ref{rhononper}) the
  $\tilde{\alpha}(T)=2.096~\alpha^{pert}(T)$ has been used, where $\alpha^{pert}(T)=\frac{g^2(T)}{4 \pi}$ (for details see \cite{Kaczmarek}).

\section{Numerical analysis}
\begin{figure}[h!]
\begin{center}
\includegraphics[width=12cm]{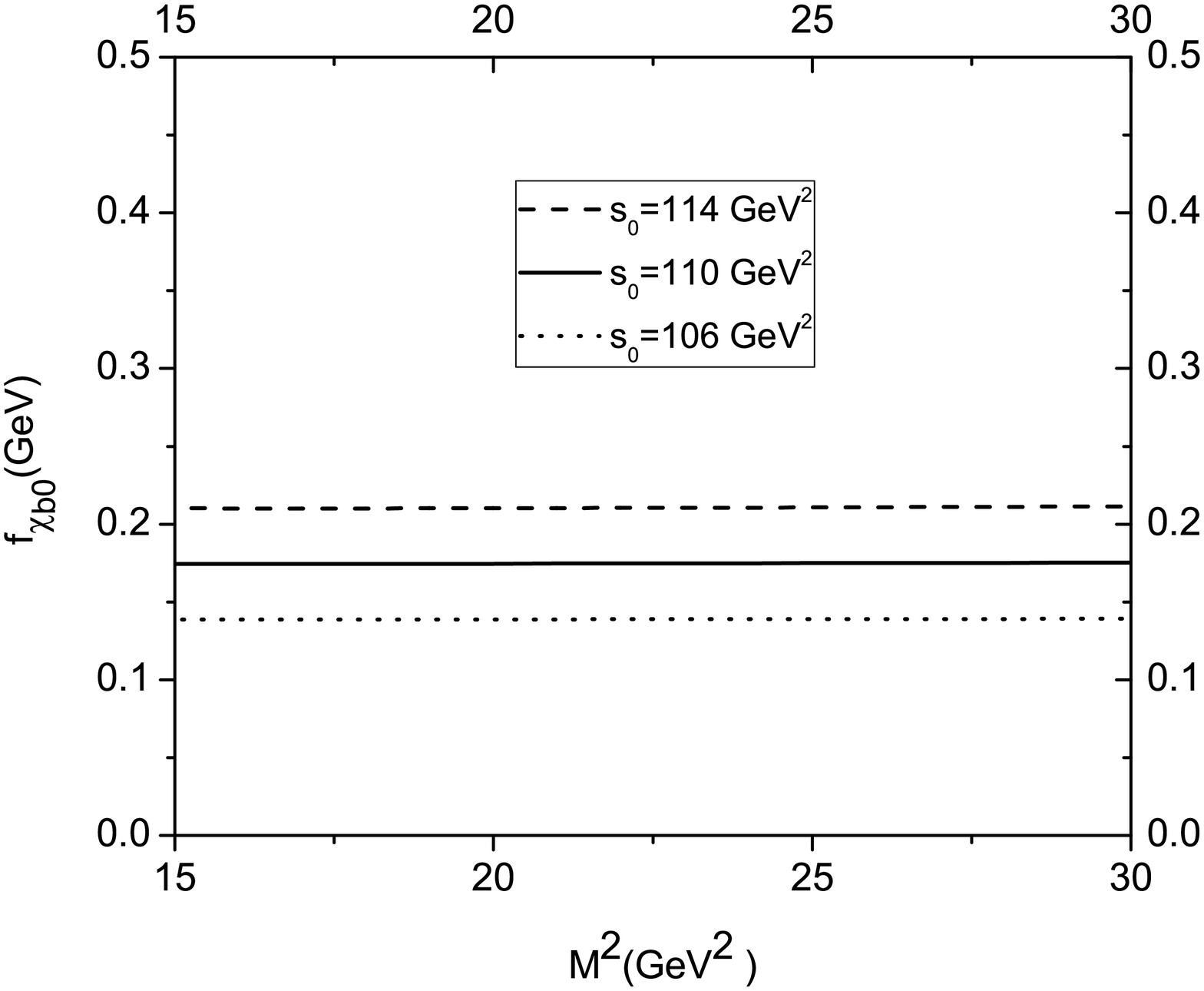}
\end{center}
\caption{The dependence of the decay constant of scalar $\chi_{b0}$ 
meson  on the Borel parameter, $M^{2}$ in vacuum at three fixed
values of the continuum threshold.} \label{fig2}
\end{figure}

\begin{figure}[h!]
\begin{center}
\includegraphics[width=12cm]{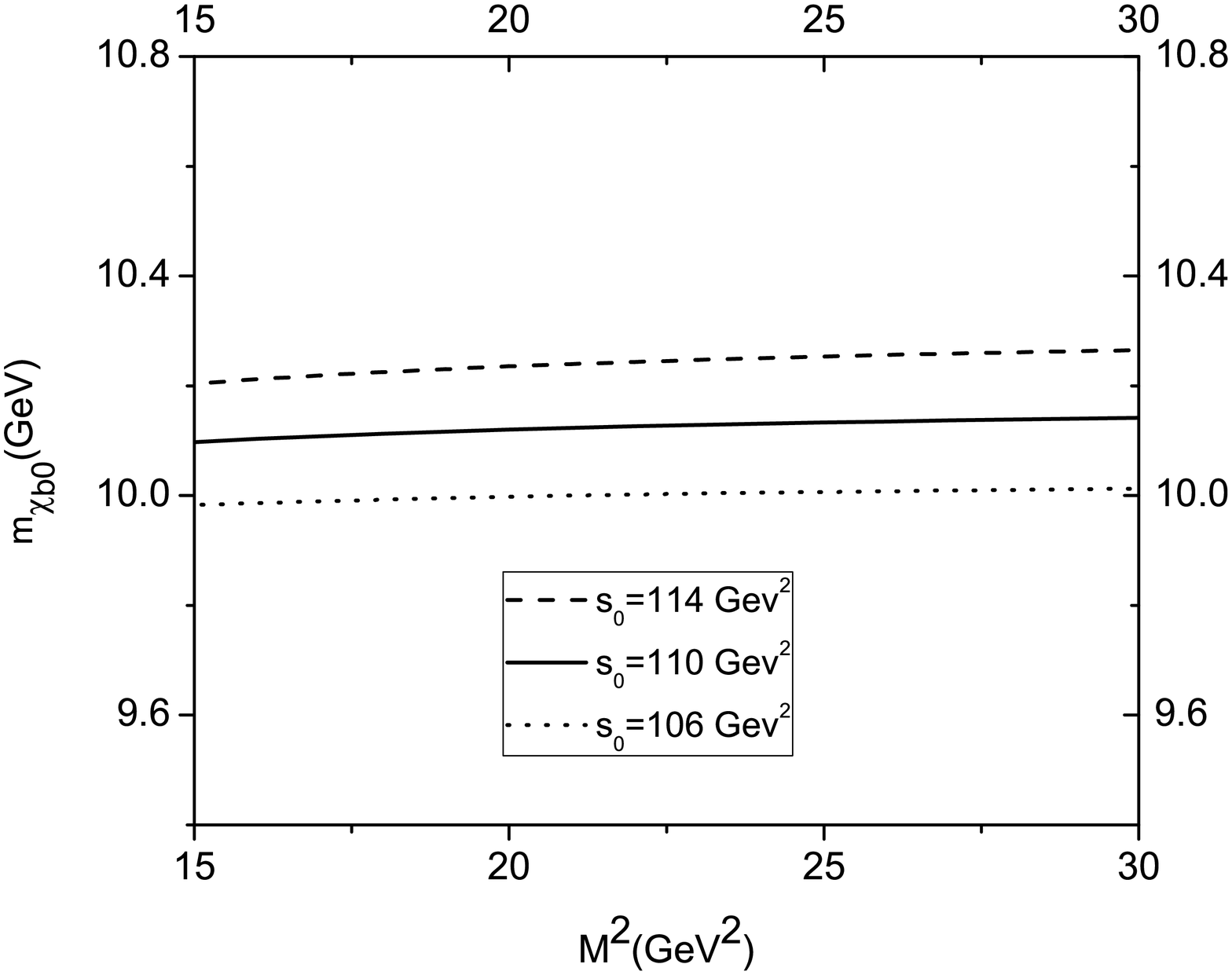}
\end{center}
\caption{The same as FIG. 2 but for mass of the scalar $\chi_{b0}$ 
meson.} \label{fig3}
\end{figure}

\begin{figure}[h!]
\begin{center}
\includegraphics[width=12cm]{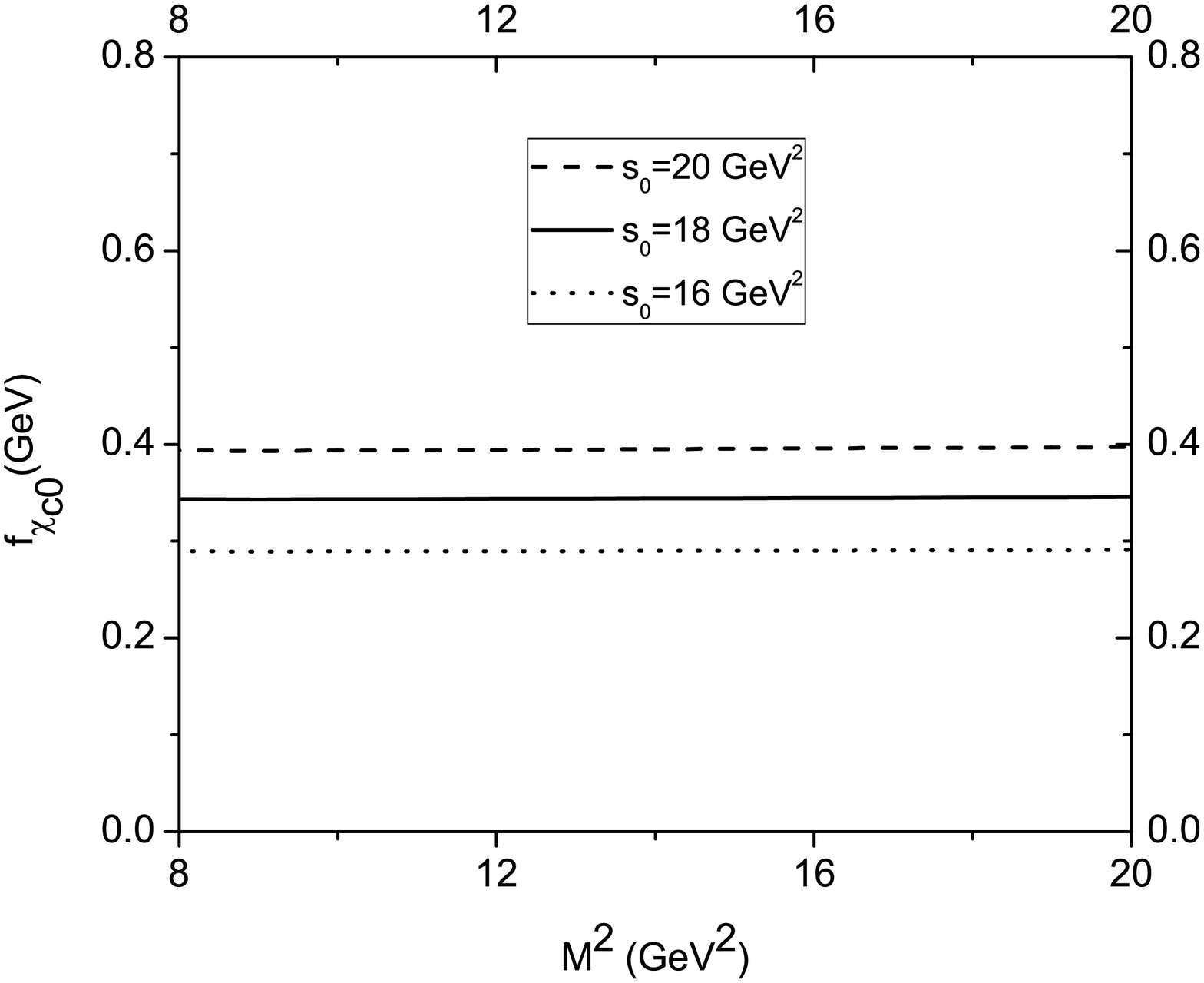}
\end{center}
\caption{The dependence of the decay constant of scalar $\chi_{c0}$ 
meson on the Borel parameter, $M^{2}$ in vacuum at three fixed values
of the continuum threshold. } \label{fig4}
\end{figure}

\begin{figure}[h!]
\begin{center}
\includegraphics[width=12cm]{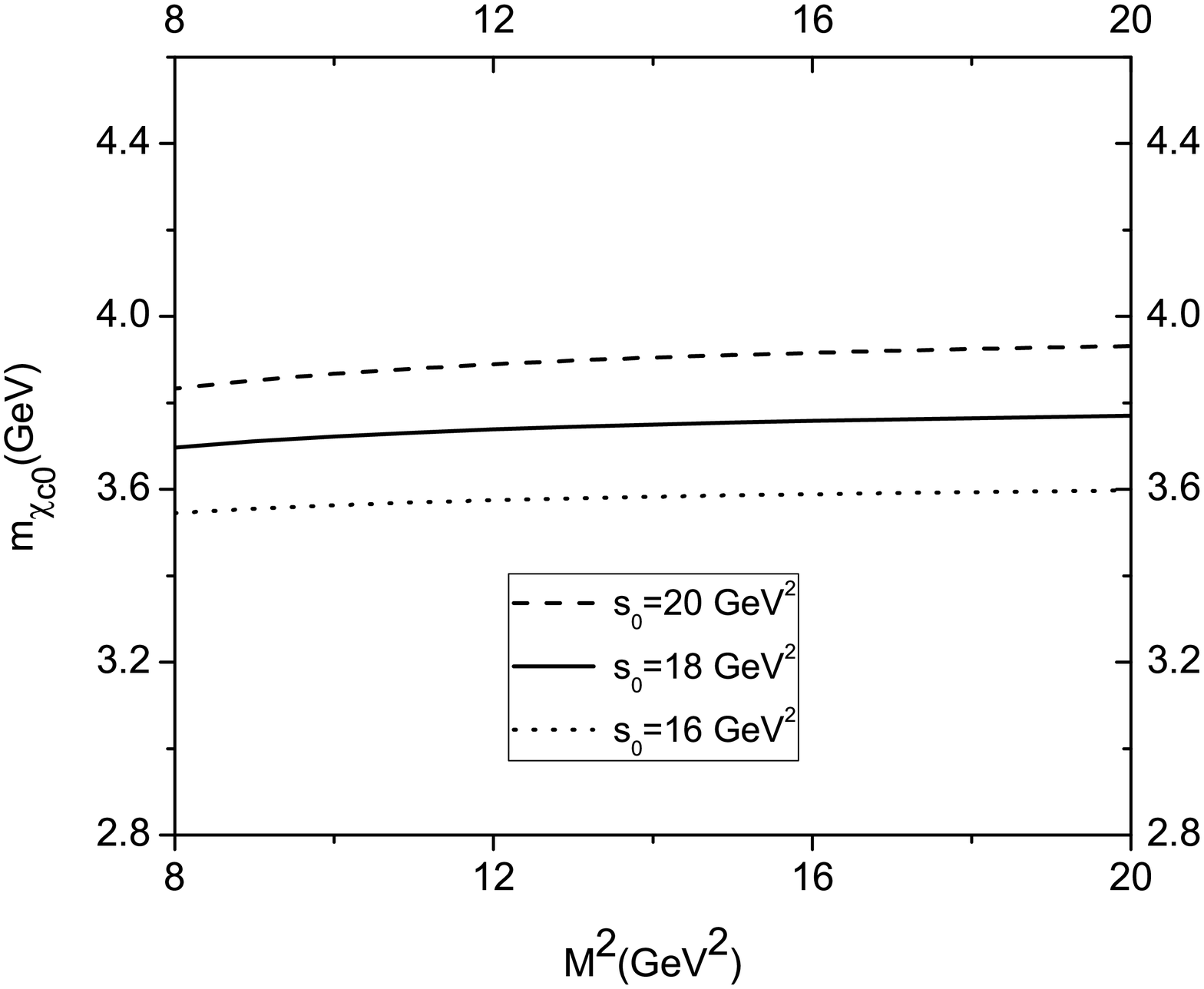}
\end{center}
\caption{The same as FIG. 4 but for mass of the scalar $\chi_{c0}$ 
meson.} \label{fig5}
\end{figure}

Present section is devoted to the numerical analysis of  the sum
rules for the  mass and decay constant of the heavy scalar mesons.
In further analysis, we use $m_c=(1.3\pm0.05)GeV$,
$m_b=(4.7\pm0.1)GeV$ and ${\langle}0\mid \frac{1}{\pi}\alpha_s G^2
\mid 0 {\rangle}=(0.012\pm0.004)GeV^4$. The sum rules  for the mass
and decay constant  also contain two auxiliary parameters, continuum
threshold $s_0$ and Borel mass parameter $M^2$.  The standard
criteria in QCD sum rules is that the physical quantities should be
independent of the  auxiliary parameters. However, the continuum
threshold $s_{0}$ is not completely arbitrary  but is  related to
the energy of the first exited state with the same quantum numbers
and can depend on the Borel mass parameter \cite{Lucha}. Therefore,
the standard criteria, does not render  realistic errors, and in
fact the existent error should be large. Hence, we will add also the
systematic errors to the numerical results. We choose the values
$s_0=(110\pm4)~GeV^2$ and  $s_0=(18\pm2)~GeV^2$  for the continuum
threshold at $\chi_{b0}$ and $\chi_{c0}$ channels, respectively. The
working region for the Borel mass parameter is determined requiring
that not only  the higher state and continuum contributions are
suppressed but also the contribution of the highest order operator
should be small, i.e., the sum rules for the mass and decay constant
should converge. As a result of the above procedure, the
working region for the Borel parameter is found to be $ 8~ GeV^2
\leq M^2 \leq 20~ GeV^2 $ for $\chi_{c0}$ and $ 15~ GeV^2 \leq M^2
\leq 30~ GeV^2 $ for $\chi_{b0}$ mesons. The dependences of the masses and decay constants at $T=0$ on Borel mass parameter  
are shown in  Figs.
\ref{fig2}-\ref{fig5}. These figures depict that  the observables  depend very weakly on the Borel mass parameter, $M^2$ in the working regions.

\begin{table}[h]
\renewcommand{\arraystretch}{1.5}
\addtolength{\arraycolsep}{3pt}
$$
\begin{array}{|c|c|c|}
\hline \hline
         &f_{\chi_{b0}}(MeV) & f_{\chi_{c0}}(MeV)      \\
\hline
  \mbox{Present Work}        &  175\pm55   &  343\pm112 \\
\hline
  \mbox{QCD sum rules \cite{Novikov}}        & -  &  359 \\
\hline
  \mbox{Cornell potential model \cite{Eichten}}        & -  &  338\\
       \hline
  \mbox{QCD sum rules \cite{P. Colangelo}}        & -  &  510\pm40 \\
             \hline \hline
\end{array}
$$
\caption{Values of the leptonic decay constants of the heavy scalar, $\chi_{b0}$
and $\chi_{c0}$ mesons in vacuum. These results have been obtained using the values $s_{0}=110~GeV^{2}$ and $M^{2}=17~GeV^{2}$  for $\chi_{b0}$, and  $s_{0}=18~GeV^{2}$ and $M^{2}=9~GeV^{2}$
 for $\chi_{c0}$ mesons. }
\label{tab:lepdecconst}
\renewcommand{\arraystretch}{1}
\addtolength{\arraycolsep}{-1.0pt}
\end{table}
\begin{table}[h]
\renewcommand{\arraystretch}{1.5}
\addtolength{\arraycolsep}{3pt}
$$
\begin{array}{|c|c|c|}
\hline \hline
 &  m_{\chi_{b0}}~(GeV)& m_{\chi_{c0}}~(GeV) \\
\hline
  \mbox{Present Work }       &  10.10\pm1.75   &  3.71\pm0.62  \\
\hline
 \mbox{Experiment \cite{CAmsler}} &  9.85944\pm 0.00042  & 3.41475\pm 0.00031  \\
 \hline \hline
\end{array}
$$
\caption{Values of the mass of the heavy scalar, $\chi_{b0}$ and  $\chi_{c0}$
mesons in vacuum. } \label{tab:mass}
\renewcommand{\arraystretch}{1}
\addtolength{\arraycolsep}{-1.0pt}
\end{table}

\begin{figure}[h!]
\begin{center}
\includegraphics[width=12cm]{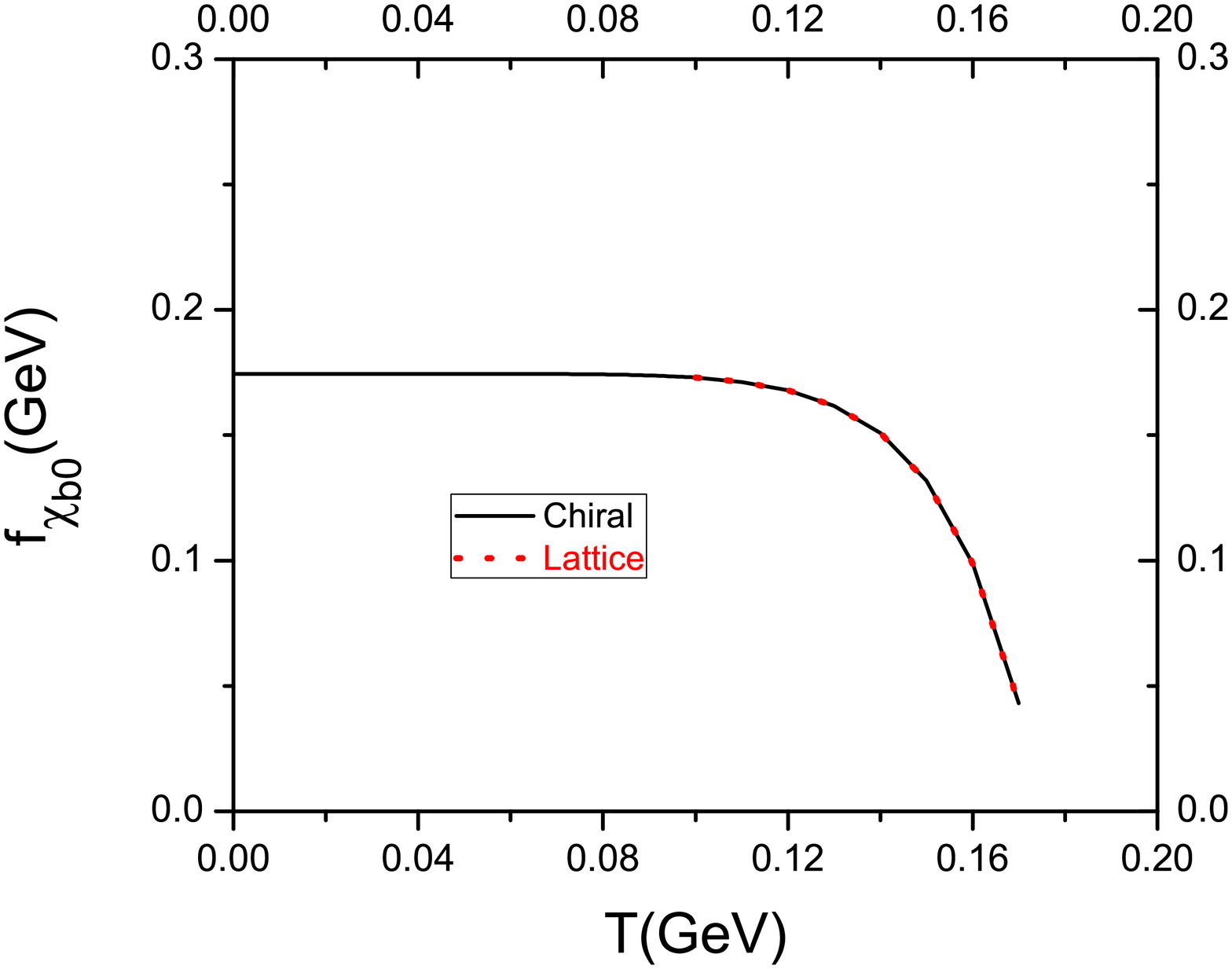}
\end{center}
\caption{The dependence of the decay constant of scalar $\chi_{b0}$ 
meson  on temperature  obtained using both lattice QCD and chiral perturbation limit values for  the gluonic part of energy density. 
 The values $s_{0}=110~GeV^{2}$, and $M^{2}=17~GeV^{2}$ have been used for the continuum
threshold and Borel mass parameter in vacuum, respectively.}
\label{mXb0Temp11}
\end{figure}

\begin{figure}[h!]
\begin{center}
\includegraphics[width=12cm]{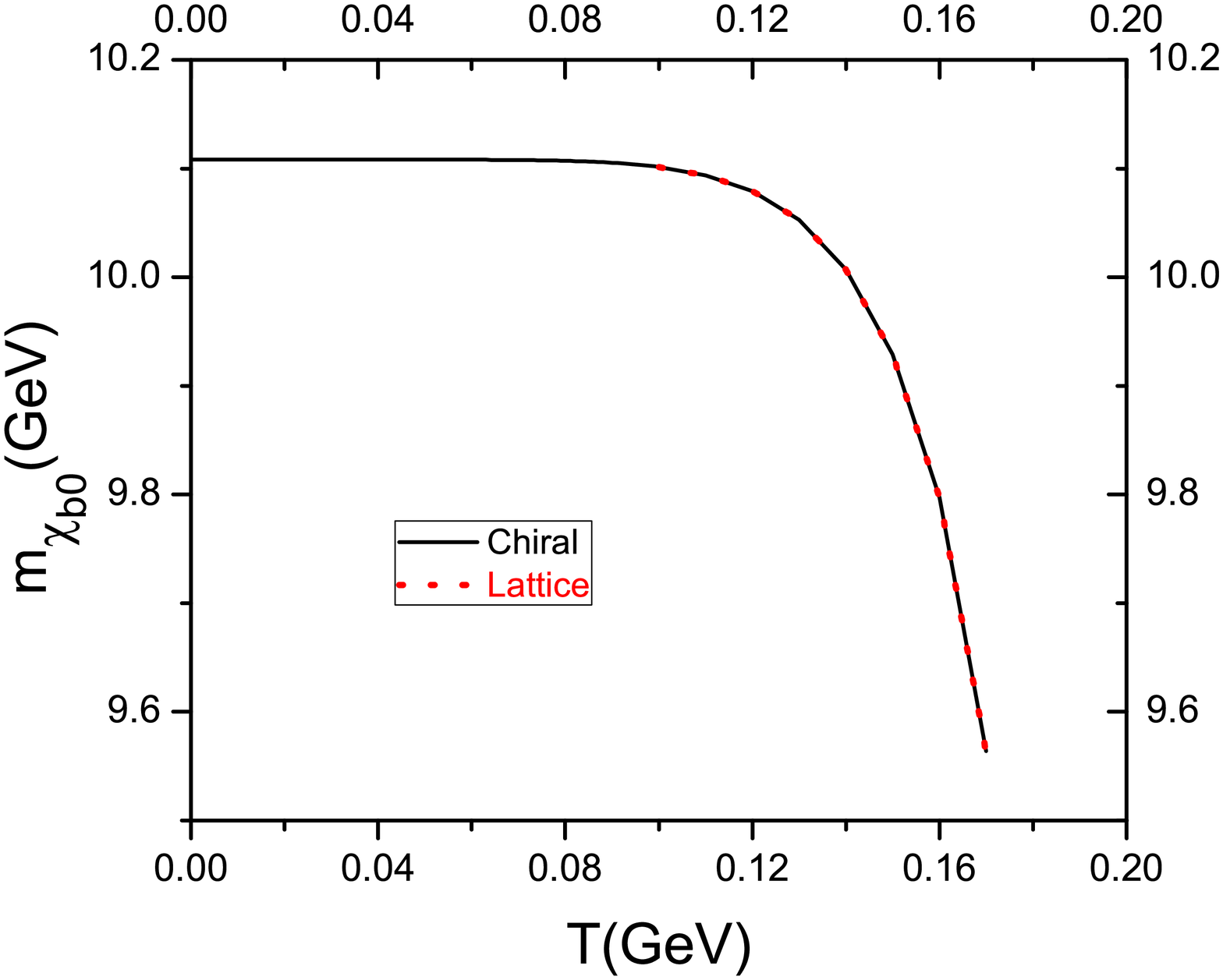}
\end{center}
\caption{The same as FIG. 6 but for mass of the scalar $\chi_{b0}$ 
meson.}
\label{mXc0Temp1}
\end{figure}

\begin{figure}[h!]
\begin{center}
\includegraphics[width=12cm]{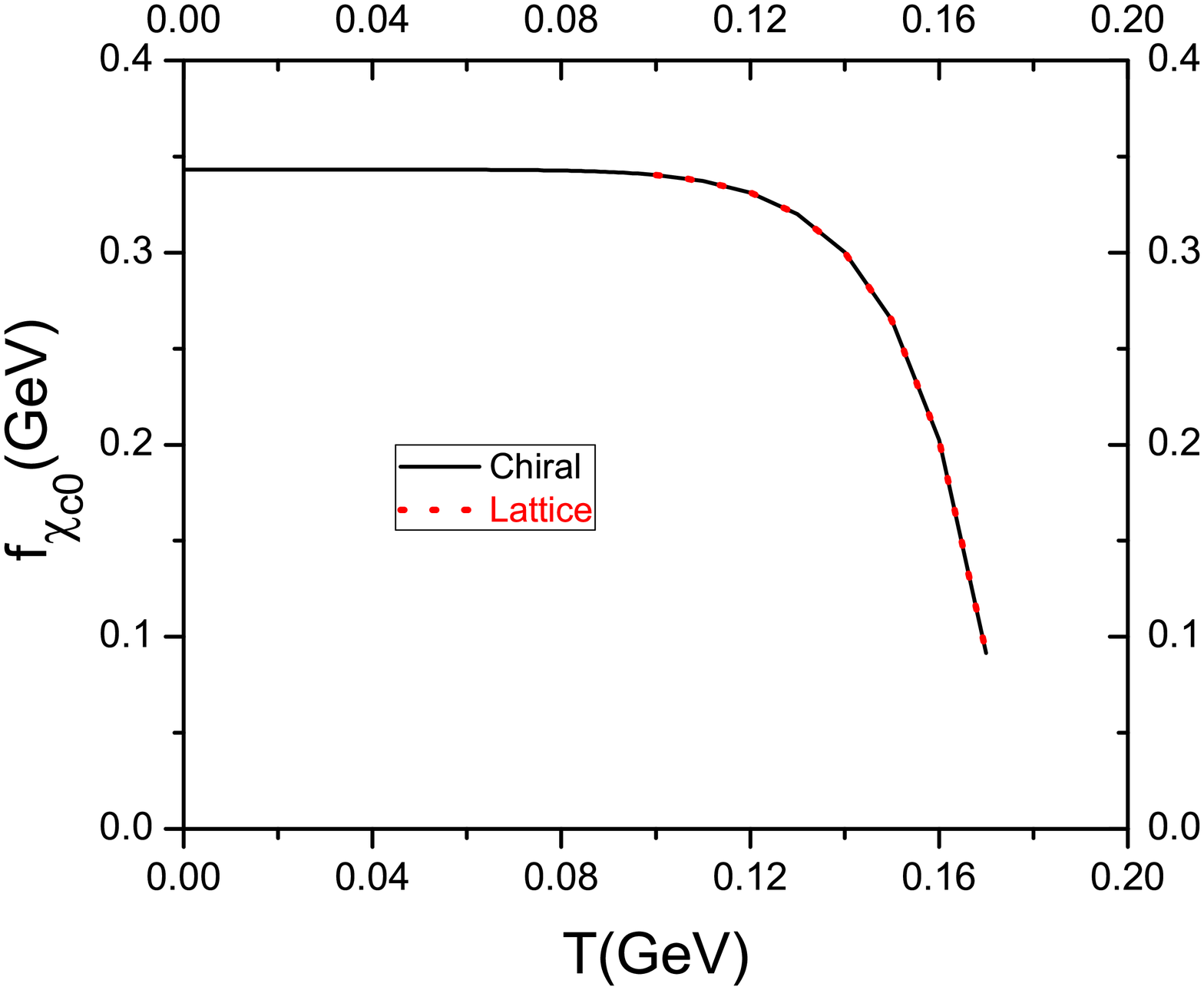}
\end{center}
\caption{The dependence of the decay constant of scalar $\chi_{c0}$ 
meson on temperature  obtained using both lattice QCD and chiral perturbation limit values for  the gluonic part of energy density. 
 The values $s_{0}=18~GeV^{2}$, and $M^{2}=9~GeV^{2}$ have been used 
 for the continuum
threshold and Borel mass parameter in vacuum, respectively.}
\label{fXb0Temp}
\end{figure}

\begin{figure}[h!]
\begin{center}
\includegraphics[width=12cm]{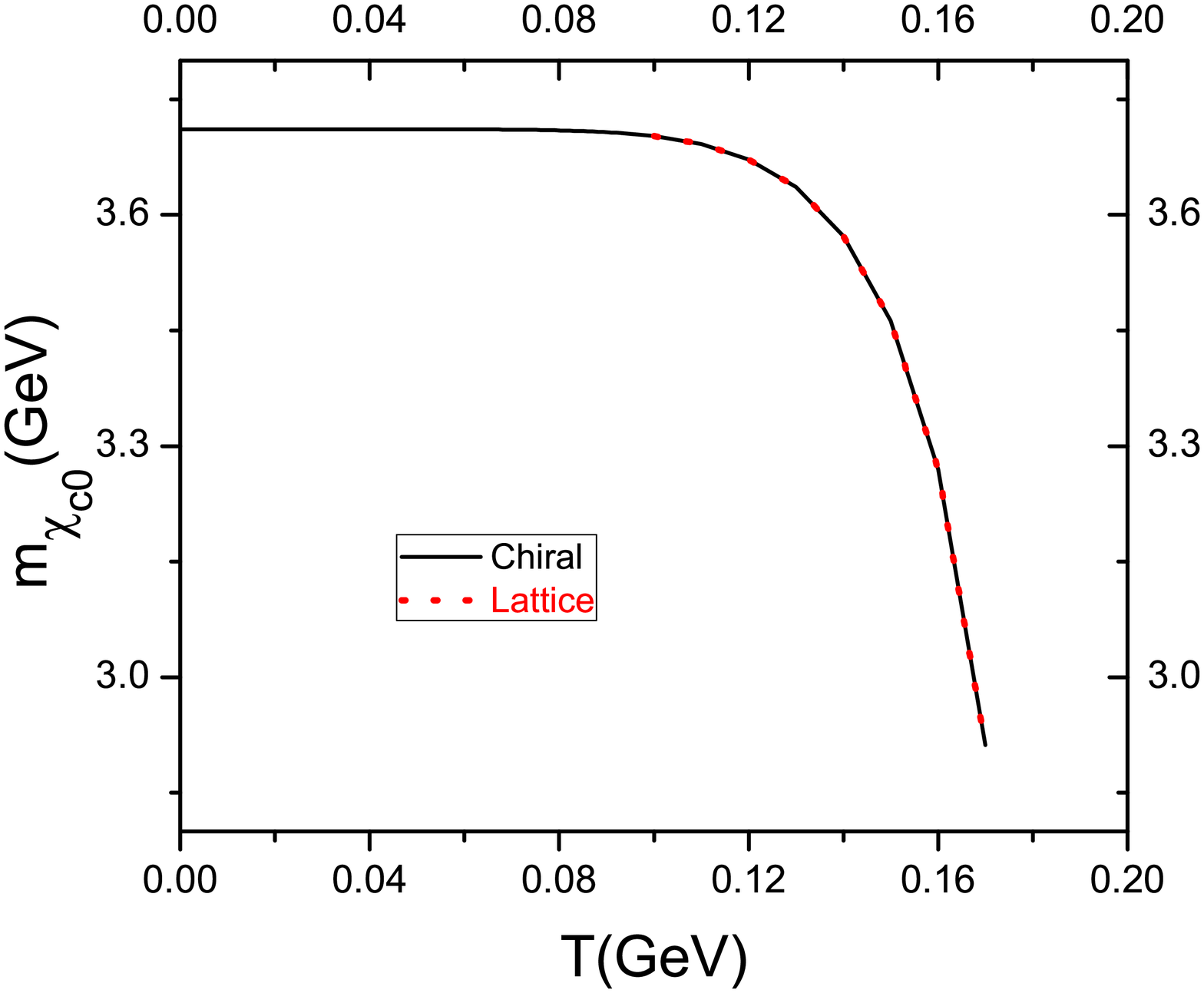}
\end{center}
\caption{The same as FIG. 8 but for mass of the scalar $\chi_{c0}$ 
meson.}
\label{mXb0Temp}
\end{figure}

The dependence of the mass
and decay constant of the $\chi_{b0}$ and $\chi_{c0}$ mesons on temperature
are presented in Figs.
\ref{mXb0Temp11}-\ref{mXb0Temp}. In these figures, we show the results obtained using both lattice QCD and chiral perturbation limit values for  the gluonic part of energy density.
These figures depict that the results depend very weakly  on the values of the gluonic part of energy density, i.e., both values obtained from lattice and chiral limit have approximately the same predictions in the  interval,
$0.1~GeV\leq T \leq 0.17~GeV$ at which the lattice results are valid. These figures also show
 that
the masses and decay constants don't change up to $T\simeq100~MeV$
 but they start to diminish with increasing the
temperature after this point. At near the critic or deconfinement temperature, the decay constants reach approximately to 25\% of their values in vacuum, while the masses are decreased about 6\% and 23\% for bottom and
charm cases, respectively.
 From these figures we deduce the results on the  decay constant and mass  in vacuum as presented in Tables I and II. The quoted errors in these Tables are due to the errors in variation of the
 continuum threshold, Borel mass parameter and errors in other input parameters
 as well as the systematic uncertainties. Table I also include a comparison of the decay constant
of charm case with the existing predictions of the same framework or other nonperturvative approaches. From this Table, we see that our predictions on the decay constant of the $\chi_{c0}$ at zero temperature
 are well consistent with the predictions of QCD sum rules \cite{Novikov} and Cornell potential model \cite{Eichten} predictions, but differ considerably from the result obtained in \cite{P. Colangelo} when the central
values are considered.
In   Table II, we also compare our predictions on the masses of the heavy scalar mesons with the existing experimental data which are in a  good consistency.
  Our results for
the leptonic decay constants as well as their behavior with respect to the temperature can be verified in the future experiments.
\section{Acknowledgment}
This work has been supported by the Scientific and Technological Research Council of  
     Turkey (TUBITAK) under research project No. 110T284.



\end{document}